\documentclass[%
 aip,
 amsmath,amssymb,
 reprint,%
]{revtex4-1}

\usepackage{graphicx}
\usepackage{dcolumn}
\usepackage{bm}
\usepackage[dvipsnames]{xcolor}
\usepackage[utf8]{inputenc}
\usepackage[T1]{fontenc}

\usepackage{mathptmx}
\usepackage{etoolbox}
\usepackage[font=small,labelfont=bf,
   justification=justified,
   format=plain]{caption}

\makeatletter
\def\@email#1#2{%
 \endgroup
 \patchcmd{\titleblock@produce}
  {\frontmatter@RRAPformat}
  {\frontmatter@RRAPformat{\produce@RRAP{*#1\href{mailto:#2}{#2}}}\frontmatter@RRAPformat}
  {}{}
}%
\makeatother
\usepackage{color}

\usepackage[colorlinks,linkcolor={}]{hyperref}
\usepackage[nameinlink,noabbrev]{cleveref}
\usepackage{movie15}
\usepackage{scalerel}
\usepackage{tikz}
\usetikzlibrary{svg.path}

\definecolor{orcidlogocol}{HTML}{A6CE39}
\tikzset{
  orcidlogo/.pic={
    \fill[orcidlogocol] svg{M256,128c0,70.7-57.3,128-128,128C57.3,256,0,198.7,0,128C0,57.3,57.3,0,128,0C198.7,0,256,57.3,256,128z};
    \fill[white] svg{M86.3,186.2H70.9V79.1h15.4v48.4V186.2z}
                 svg{M108.9,79.1h41.6c39.6,0,57,28.3,57,53.6c0,27.5-21.5,53.6-56.8,53.6h-41.8V79.1z M124.3,172.4h24.5c34.9,0,42.9-26.5,42.9-39.7c0-21.5-13.7-39.7-43.7-39.7h-23.7V172.4z}
                 svg{M88.7,56.8c0,5.5-4.5,10.1-10.1,10.1c-5.6,0-10.1-4.6-10.1-10.1c0-5.6,4.5-10.1,10.1-10.1C84.2,46.7,88.7,51.3,88.7,56.8z};
  }
}

\newcommand\orcidicon[1]{\href{https://orcid.org/#1}{\mbox{\scalerel*{
\begin{tikzpicture}[yscale=-1,transform shape]
\pic{orcidlogo};
\end{tikzpicture}
}{|}}}}

\usepackage{hyperref}
\usepackage{subcaption}
\usepackage{caption}
\usepackage{graphicx}
\begin{document}

\preprint{AIP/123-QED}

\title[Generation of Strong Fields \textcolor{black}{with Subcritical Density Plasmas} to Study the Phase Transitions of Magnetized Warm Dense Matter]{Generation of Strong Fields \textcolor{black}{with Subcritical Density Plasmas} to Study the Phase Transitions of Magnetized Warm Dense Matter}
\author{\orcidicon{0000-0002-7255-6652} I. N. Erez}
\affiliation{Extreme State Physics Laboratory, Physics and Astronomy Department, University of Rochester, Rochester, New York 14627, USA}
\affiliation{Laboratory for Laser Energetics, University of Rochester, 250 E. River Road, Rochester, New York 14623, USA}
\email{ierez@lle.rochester.edu}

\author{\orcidicon{0000-0001-8233-8272} J. R. Davies}%
\affiliation{Laboratory for Laser Energetics, University of Rochester, 250 E. River Road, Rochester, New York 14623, USA}

\author{\orcidicon{0000-0001-6488-3277} J. L. Peebles}%
\affiliation{Laboratory for Laser Energetics, University of Rochester, 250 E. River Road, Rochester, New York 14623, USA}

\author{\orcidicon{0000-0001-8742-0304} R. Betti}%
\affiliation{Laboratory for Laser Energetics, University of Rochester, 250 E. River Road, Rochester, New York 14623, USA}

\author{\orcidicon{0000-0002-3272-2193} P.-A. Gourdain}
\affiliation{Extreme State Physics Laboratory, Physics and Astronomy Department, University of Rochester, Rochester, New York 14627, USA}
\affiliation{Laboratory for Laser Energetics, University of Rochester, 250 E. River Road, Rochester, New York 14623, USA}

\date{\today}


\begin{abstract} 


Warm dense matter (WDM) is a regime where Fermi degenerate electrons play an important role in the macroscopic properties of a material. Recent experiments have brought us closer to understanding unmagnetized processes in WDM, but magnetized WDM remains unexplored because \textcolor{black}{kilotesla} magnetic fields are required. Although there are examples of field compression generating such fields by \textcolor{black}{imploding} pre-magnetized targets, these existing methods give no independent control over the parameters of the magnetized plasma and result in limited \textcolor{black}{laser access for sample creation and diagnosis}. In this paper, numerical simulations show that \textcolor{black}{kilotesla} magnetic fields can be obtained by shining laser beams onto the inner surface of a cylindrical target, rather than on the outer surface. This approach relies on field compression by a low-density high-temperature plasma, rather than a high-density, low-temperature plasma, used in the more conventional approach. With this novel configuration, the region of peak magnetic field is mostly free of plasma, hence \textcolor{black}{other} beams can reach a sample placed in the region of the peak field \textcolor{black}{to form WDM and diagnose it}.

\end{abstract}

\maketitle

\section{\label{sec:intro}Introduction}


Warm dense matter (WDM) is defined as strongly-coupled plasma,\cite{RevModPhys.54.1017} with temperatures in the eV range and densities above solid density \cite{Graziani2014FrontiersAC}. \textcolor{black}{Initial studies} date back to the 1980's \cite{RevModPhys.54.1017}. WDM is a relatively unexplored state of matter lying between condensed matter and plasma regimes \cite{Graziani2014FrontiersAC}. It is relevant to fusion science \cite{doi:10.1063/1.5018580} and to a diverse group of astrophysical objects, including exoplanets and white dwarfs \cite{Koenig_2005, doi:10.1063/1.3116505, Benuzzi-Mounaix_2014, vailionis_evidence_2011, remington_accessing_2005}.
In the WDM regime, Fermi degenerate electrons play an important role in the macroscopic properties of the material.

Theoretical work on WDM is expanding thanks to Kohn–Sham density functional theory (DFT) \cite{PhysRevE.52.6202} and ab-initio simulations involving Path Integral Monte Carlo (PIMC) methods \cite{bonitz_ab_2020,dornheim_ab_2018,dornheim_uniform_2018,hamann_dynamic_2020,Gericke_2010, PhysRevLett.98.190602, sim, PhysRevE.103.033204} \textcolor{black}{for modeling} the inner structure of WDM \textcolor{black}{or} investigating thermal behavior. The significant effort in the theory community needs corresponding experiments to be performed and have WDM samples diagnosed under various conditions.
However, creating WDM in the laboratory is not a trivial task. One way to reach these high energy densities is to use lasers. Ultrafast laser-induced microexplosions confined inside a sapphire have been used for tabletop studies of WDM \cite{vailionis_evidence_2011}. Intense soft X-ray photoionization is another way to create WDM \cite{nagler_turning_2009}. Not only is the creation of WDM in the laboratory non-trivial, but probing it is also difficult, requiring methods such as inelastic X-ray scattering\cite{garcia_saiz_probing_2008}.

Studying phase transitions is of critical importance for characterizing any state of matter. Using diamond-anvil-cells (DACs) and shock-wave experiments, the equation of state (EOS) of materials at high pressures has been studied\cite{lomonosov_multi-phase_2007}. 
There are also studies on melting curves of various metals at high pressures,\cite{parisiades_review_2021, lord_melting_2014} and this is a growing field of research.

One unexplored interest in the WDM regime is the effect of electron magnetization \textcolor{black}{on the material properties, such as phase transitions}. A rare example presents \textcolor{black}{Particle-In-Cell (PIC)} simulations demonstrating the benefit of magnetic fields for the creation of uniform WDM samples \cite{PhysRevE.101.051202}. Our interest in magnetized WDM and its phase transitions is not only fundamental science, but is also helpful for magneto-inertial fusion \cite{doi:10.1063/5.0091529,osti_249559,doi:10.1063/1.3333505,doi:10.1063/1.4920948,6353648}, \textcolor{black}{and astrophysics} \cite{RevModPhys.54.1017, RevModPhys.78.755}. However, due to technical limitations\textcolor{black}{,} magnetized WDM has yet to be \textcolor{black}{produced} experimentally.

First, the field must be compressed to reach strengths capable of magnetizing the highly collisional electrons. To estimate the minimum magnetic field \textcolor{black}{strength} required, the collision frequency in WDM $\nu_{col,WDM}$ should be less than the electron cyclotron frequency $\omega_{ce}$. \textcite{meyer-ter-vehn_collisional_2019} calculate the effective dynamic collision frequency for WDM while deriving the rate of photon absorption induced by free-free electron-ion collisions. Referring to their work\cite{meyer-ter-vehn_collisional_2019} with common WDM sample parameters seen in OMEGA experiments\cite{PhysRevLett.112.155003, Lahmann_2023, fletcher_x-ray_2013, PhysRevB.86.144115}, so the ionization state of 1, temperatures in the range of $0.1$ eV to $100$ eV, photon energy corresponding to an OMEGA beam of wavelength $351$ nm, and at densities $0.1$ to $10$ times the solid density, the collision frequency would be in the range of $10^{14}\,s^{-1}$ to $10^{17}\,s^{-1}$\cite{meyer-ter-vehn_collisional_2019}. Even though the theory presented in \textcite{meyer-ter-vehn_collisional_2019} would not be sufficient to explain WDM densities larger than $0.1$ times the solid density, it can still provide a lower theoretical limit for the minimum magnetic field strength required for ten percent of the solid density. In fact, for a temperature of 100 eV, the minimum field strength required would be 0.5 kT while for 10 eV, it would be 5 kT. Regarding the \textcolor{black}{restricted} \textcolor{black}{knowledge of the subject} and the laser energy for 351 nm being in the \textcolor{black}{bounds} of the theory, as an order of magnitude estimate, that lower limit for the magnetic field strength can be taken to be 1 kT. We consider this lower limit as a basis since the setup we present in this manuscript is generalizable to various sample types. Second, the high-power beam required for isochoric heating \textcolor{black}{of the sample to create the WDM state} should reach the sample surrounded by the flux compression plasma\cite{PhysRevLett.91.125004}. Finally, the samples must be diagnosed, typically using X-rays, which need to traverse the compression plasma and the sample before being measured. This requires a setup capable of accommodating the flux compression geometry and facilitating field measurements to relate changes in material properties \textcolor{black}{caused by} magnetization.

Kilotesla magnetic fields were reached more than a decade ago, using laser-driven magnetic field compression\cite{doi:10.1063/1.3416557}. More recent work shows that laser-driven cylindrical implosions can compress initial fields of $5\,\text{T}$ to $25\,\text{kT}$ for magnetized implosions\cite{PhysRevE.106.035206}. Other research showed that kilotesla magnetic field strengths via laser-driven flux compression were possible in a lab environment \cite{PhysRevLett.103.215004,doi:10.1063/1.3696032}. However, the challenge here is producing large magnetic fields while getting the necessary access to create and diagnose WDM \textcolor{black}{along with} measuring the field. These requirements rule out immediately the most common techniques used by the high energy density plasma community, which relies on cylindrical flux compression to generate large magnetic fields\cite{doi:10.1063/1.3416557, PhysRevE.106.035206, PhysRevLett.103.215004,doi:10.1063/1.3696032}. In these setups, the cylindrical compression comes from a high-density plasma converging radially, defacto increasing the field. However, the convergent geometry closes all access to the sample, except axially. Yet \textcolor{black}{the axial direction cannot incorporate the heating beam for sample creation, the diagnostic for phase transitions, and the tool for initial magnetic field strength creation simultaneously. In that case, it would be necessary to have the sample on an open end of the cylinder along the cylinder axis to avoid heating beam passing through high-density plasma. This would make it impossible to allocate space for sample diagnosis, which requires X-ray backlighting of the sample. Another important concern would be that it was proved difficult to measure the compressed magnetic field in the converging design with proton radiography \cite{doi:10.1063/1.1788893, peebles_axial_2020, heuer_diagnosing_2022}.} 

This paper explores how magnetic flux compression can be achieved using a low-density plasma that would facilitate the formation and diagnosis of magnetized WDM. This is accomplished by heating a low-density plasma to reach high pressures. To achieve this\textcolor{pink}{,} we propose a setup involving half a hohlraum \cite{doi:10.1063/1.871209} with an initial axial magnetic field of $50\,\text{T}$ where the low-density plasma ablated from the inner surface of the halfraum compresses the magnetic field. This proposed setup would allow the heating beam to reach the sample and create WDM while maximizing \textcolor{black}{X}-ray transmission. In the rest of the paper, section \ref{sec:design} presents the proposed target design which \textcolor{black}{is} called "halfraum". Then, the results of the 2-D magnetohydrodynamic simulation in PERSEUS\cite{doi:10.1063/1.3543799} are shared in section \ref{sec:sim}. The findings are discussed in section \ref{sec:discussion}.


\section{\label{sec:design}Design}

Cylindrical implosions with high-power lasers are used routinely to compress magnetic fields  \cite{PhysRevLett.103.215004, osti_10300809}, even reaching kilotesla strengths. These laser-driven magnetic field compression setups usually involve a convergent geometry with initially applied field lines being compressed with laser beams pointed to the outer surface of the target. These setups benefit from \textcolor{black}{high-density} plasma created with laser ablation to compress the magnetic field lines. However, the converging configuration blocks the high field region from direct laser illumination, a necessary requirement to produce WDM, at least along the radial direction. While the axial direction can be used, it then becomes difficult to simultaneously heat the sample, diagnose it, and measure the field using \textcolor{black}{proton radiography}. Therefore, the conventional compression approach is less than ideal not only for producing WDM but also for diagnosing it while measuring the field. 

\begin{figure*}[!htb]
    \centering
    \begin{subfigure}[t]{0.4\textwidth}
        \centering
        \includegraphics[width=\textwidth]{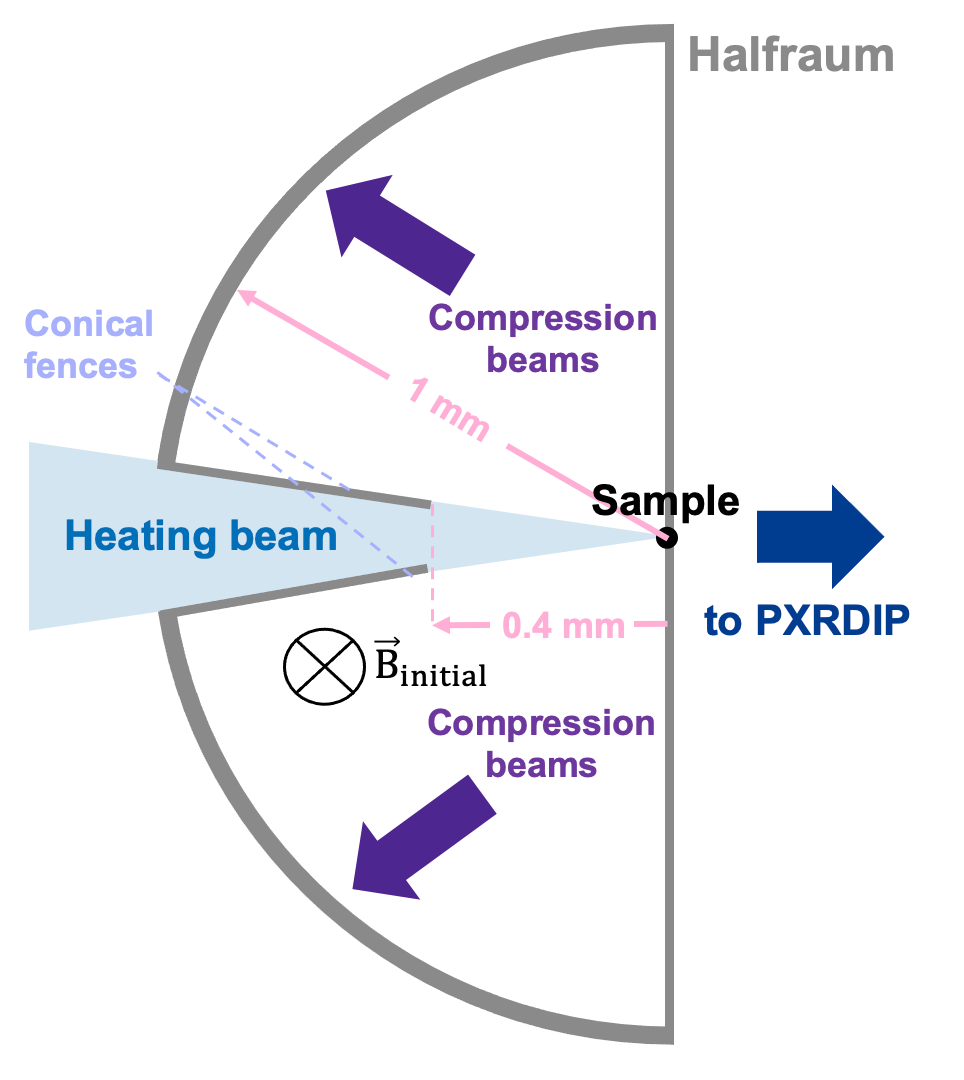}
        \caption{Polar view of the "halfraum" setup.} \label{fig:setup}
    \end{subfigure}%
    ~ 
    \begin{subfigure}[t]{0.6\textwidth}
        \centering
        \includegraphics[width=\textwidth]{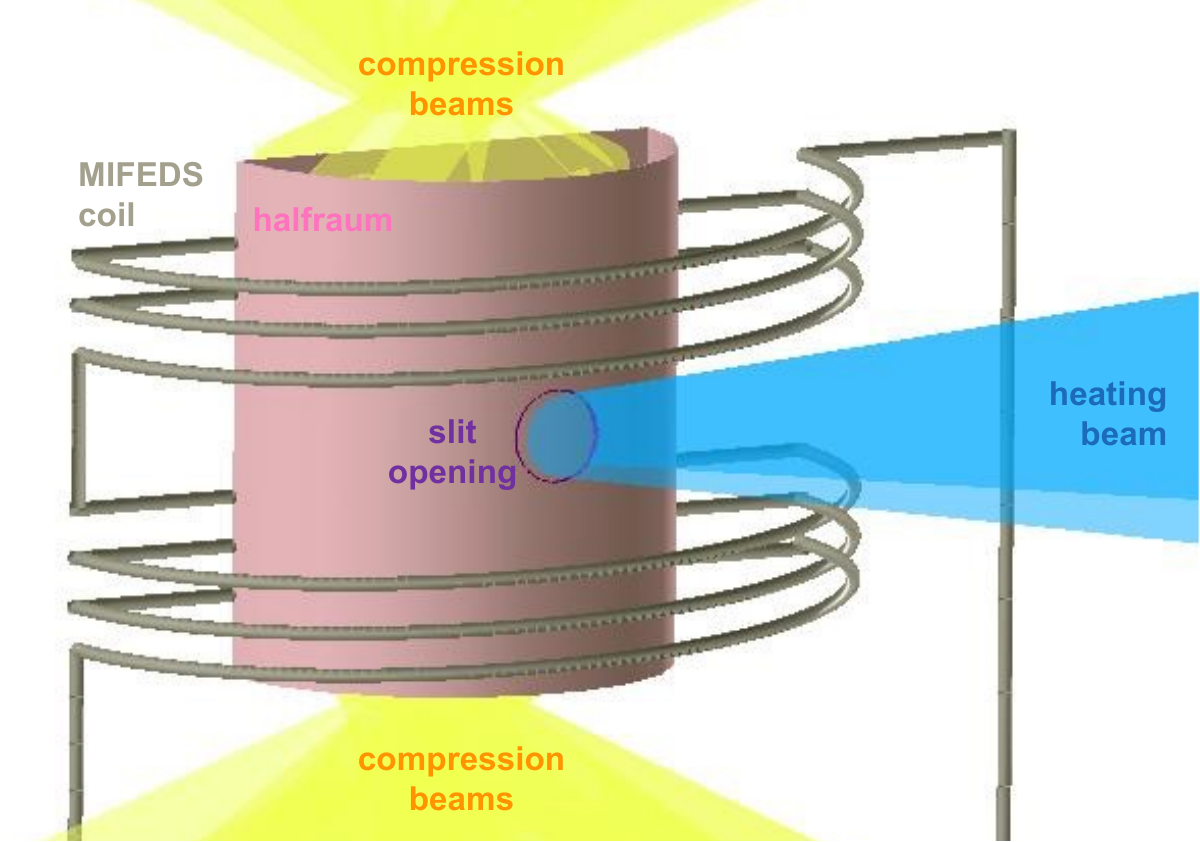}
        \caption{VisRad model of the setup presented.}\label{fig:app1} 
    \end{subfigure}
    \caption{(a) Polar view of the "halfraum" setup refers to a hohlraum cut along its axial plane to form a D-shaped prism. The half geometry is needed to allow space for diagnostic hardware, such as PXRDIP, to record the phase transition data. The halfraum is ablated from the inside with polar drive beams to compress the axial magnetic field lines initially applied by MIFEDS 2.5 \cite{mifeds} \textcolor{black}{in order} to create \textcolor{black}{kilotesla} order magnetization at the sample location. The setup involves a slit with a conical entrance to enable the heating beam to reach the sample for the creation of WDM state. (b) VisRad model of the setup presented provides a 3D view of the halfraum setup. Here, the opening for the heating beam and the MIFEDS coil wrapped around the halfraum can be better seen.}
\end{figure*}

In order to allow for the sample to be less affected by the field compression, we propose redirecting the compression beams to the inner surface of \textcolor{black}{a} cylinder, in a polar drive arrangement reminiscent of the hohlraum used in fusion experiments at the \textcolor{black}{National Ignition Facility (NIF)}\cite{doi:10.1063/1.871209}. The cross-section along the radial plane is shown in \autoref{fig:setup}. Using this setup, magnetic flux can be compressed by the low-density plasma ablated by the polar drive beams while keeping the sample in an environment accessible to the heating beam. However, the sample also needs to be diagnosed\textcolor{black}{. For example, the Powder X-Ray Diffraction Image Plate (PXRDIP)\cite{pxrdip} diagnostic is used for an interest in phase transitions}, which utilizes \textcolor{black}{X}-ray Bragg scattering to track the phase transition \textcolor{black}{inside the sample}. \textcolor{black}{This diagnostic requires a flat surface and space for the PXRDIP box. For a cylindrical geometry with the sample along the axis, this translates to a 1/2-hohlraum design}, subsequently named "halfraum." \textcolor{black}{Here,} the halfraum refers to a hohlraum cut along its axial plane forming a cylinder with a D-shaped cross-section. Note that traditionally \cite{MACLAREN2006398,JOHNS2021100939}, a halfraum is a hohlraum cut along the polar plane to form a much shorter hohlraum which is different from the halfraum design presented here. Another difference is the conical opening\textcolor{black}{. This opening} was included to \textcolor{black}{reduce the plasma density in the vicinity of the WDM sample.} While this work focuses on presenting a design to study the phase transitions of magnetized WDM, it is applicable to \textcolor{black}{other cases where strong magnetization and diagnostic access are required.} Therefore, the principle behind the presented design can be used to study any magnetized \textcolor{black}{high-energy-density (HED)} matter experimentally.

\begin{figure}[!htb]
\centering  
\includegraphics[width=0.5\textwidth]{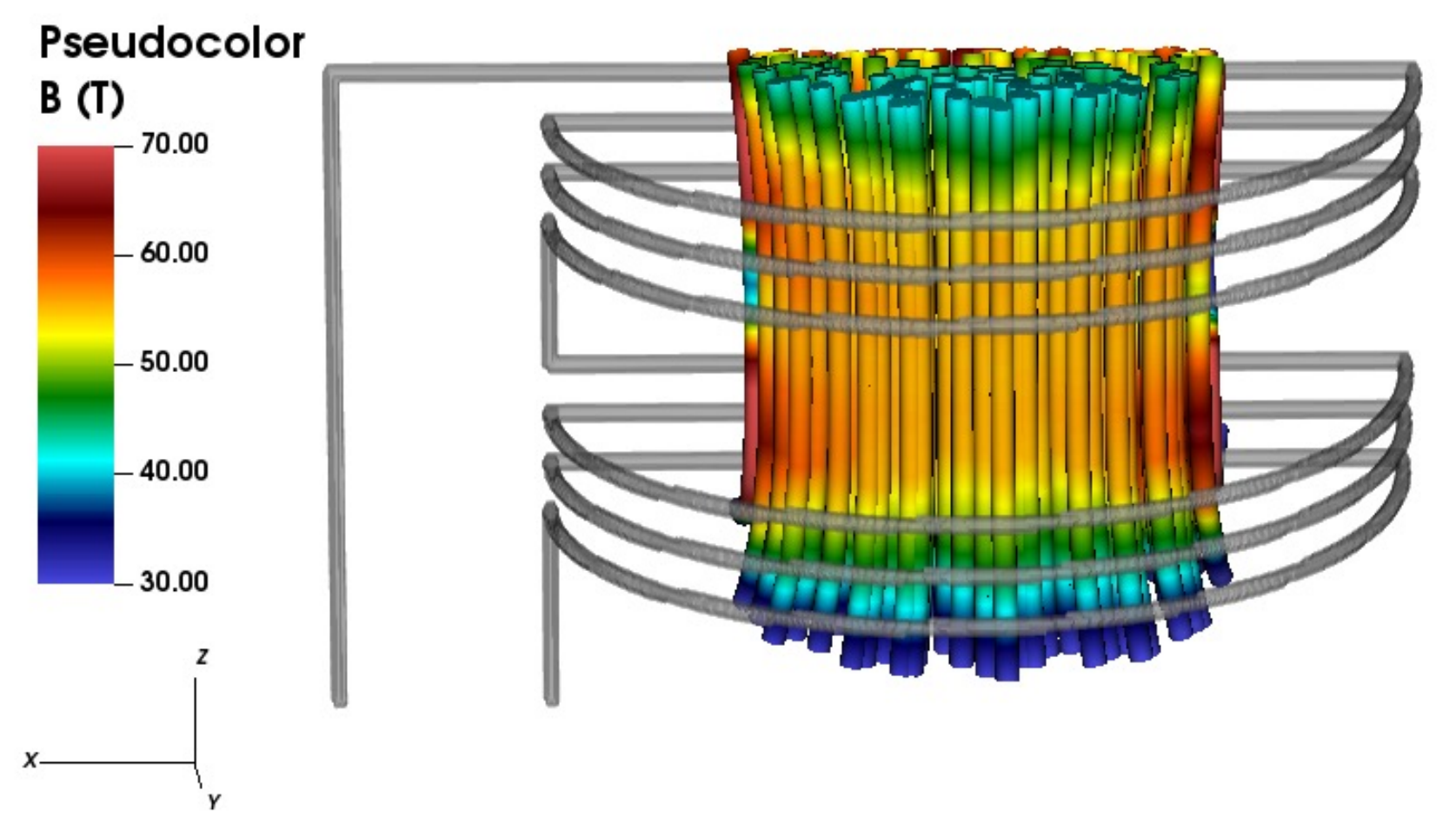}
\caption{The coil geometry design \textcolor{black}{coupled with MIFEDS 2.5 \cite{mifeds}} introduces the initial axial magnetic field lines. The half-solenoid-shaped coil with a radius of $1.75\,\text{mm}$ is wrapped around the halfraum. MIFEDS 2.5 can provide a peak current of $25\,\text{kA}$ to this coil setup of inductance $85.9\,\text{nH}$ so that the pictured magnetic fields inside the halfraum can reach the strength $50$ T.} 
\label{fig:Blines} 
   \end{figure}

\autoref{fig:setup} shows a cross-section of the proposed setup, with the sample located on the cylinder axis. It should be noted that \autoref{fig:setup} is not a scaled figure and it is a diagram. \autoref{fig:app1} shows the \textcolor{black}{VisRad\cite{visrad}} model of the setup so \textcolor{black}{the Magneto-Inertial Fusion Electrical Discharge System (MIFEDS) 2.5\cite{mifeds} coil wrapped around the halfraum to provide initial magnetic field} can be seen. The height of the halfraum is $2.2$ mm. For our purposes, a halfraum of inner radius $1\,\text{mm}$ with a thickness of \textcolor{black}{$8$} \textmu m is \textcolor{black}{used}. The setup presented here is optimized for the OMEGA laser\textcolor{black}{,} but can be \textcolor{black}{scaled} easily to other laser facilities such as NIF or \textcolor{black}{Laser Megajoule (LMJ)}. The halfraum material is plastic and simulated for carbon with an ionization state of 4 and this ionization state is assumed to stay the same throughout the simulation. An \textcolor{black}{initial} axial magnetic field of $50\,\text{T}$ is applied externally using MIFEDS 2.5. While generating such large fields with MIFEDS has not been reported in the literature, this is mostly because MIFEDS is usually used in conjunction \textcolor{black}{with coils placed far from the target to allow the beams to reach the outer surface of the cylinder.} However, our polar drive arrangement does not require access to the outer cylinder surface. Therefore, the coils can be wrapped close to the halfraum \textcolor{black}{allowing to reach} $50\,\text{T}$ \textcolor{black}{magnetic field strength as shown} in \autoref{fig:Blines}. \textcolor{black}{Another advantage of the design is that the MIFEDS capacitors would need to store $10.5$ J of energy to be converted to magnetic energy to provide an initial magnetic field of $50$ T. This is much less than the energy storage capabilities of MIFEDS 2 with $250$ J and MIFEDS 3 with $2.25$ kJ.}

Since the PXRDIP needs to be close to the WDM sample, we use a 6-turn solenoid with a D-shaped cross-section, similar to the one of the halfraum. We use a radius of $1.75\,\text{mm}$ for the half-solenoid, and its geometry is shown in \autoref{fig:Blines}. The magnetic field strengths shown in \autoref{fig:Blines} are calculated using Biot-Savart law for the half-solenoid coil. Note that the solenoid was rendered with some transparency to reveal the magnetic fields within it. The magnetic field lines in the figure are only shown in the volume circumscribed inside the halfraum. The coil has a wire radius of $30$ \textmu m, with $0.15\,\text{mm}$ spacing between the turns, and a spacing of height $0.6\,\text{mm}$ in the middle, between the third and fourth \textcolor{black}{coil turns}. \textcolor{black}{This allows} the heating beam to pass through the slit opening \textcolor{black}{of} the conical fence \textcolor{black}{as in \autoref{fig:app1}. The base radius of the conical fence} \textcolor{black}{is} $0.2\,\text{mm}$ \textcolor{black}{and it is} attached to the halfraum. That conical fence is cut at a distance of $0.4\,\text{mm}$ away from the sample as can be seen in \autoref{fig:setup}. For our proposed geometry, the coil inductance comes to $85.9\,\text{nH}$, yielding a peak \textcolor{black}{MIFEDS} current of $25\,\text{kA}$\textcolor{black}{, which} \textcolor{black}{can yield} the initial magnetic field of $50\,\text{T}$ within the halfraum \textcolor{black}{(see \autoref{fig:Blines})}. 

\textcolor{black}{Before proceeding with the simulation to test the design, it is worth mentioning the plan for a full experimental campaign along with the measurements for the magnetic field strength. While the design presented here is too complex to incorporate field compression measurements, such an experimental campaign would need to involve field compression measurements before proceeding with the halfraum design. For that former step, proton radiography could be the main diagnostic along the axis of the halfraum design to measure the compressed field strength at various times to evaluate the agreement with the simulations. In that case, synthetic proton radiographs can be generated based on simulation outputs to compare to the experimental results. Once the field strength is measured and confirmed to be in agreement with the simulations, then the setup presented here could be used to proceed.}

\section{\label{sec:sim}Simulation}

The novelty of our setup lies in \textcolor{black}{the} geometry and the way it \textcolor{black}{allows other} beams to create and diagnose a \textcolor{black}{magnetized} WDM sample. The \textcolor{black}{simulation} presented in this section ultimately try to answer \textcolor{black}{one main question}: Can the design presented here compress the magnetic field to more than $1$ kT while allowing the heating beam to penetrate through the ablated plasma to reach the sample for the creation of magnetized WDM? The 2-D \textcolor{black}{magnetohydrodynamics (MHD)} code PERSEUS\cite{doi:10.1063/1.3543799} is used to answer \textcolor{black}{these questions}. MHD can capture the behavior of the plasma and its interaction with the field to check the feasibility of the proposed setup. PERSEUS is an extended MHD code, meaning that it includes the generalized Ohm's law. This enables PERSEUS to capture resistivity and laser-plasma interactions in a non-linear regime leading to realistic yet computationally feasible solutions. However, as 2-D MHD simulations do not capture three-dimensional instabilities leading to possibly lower convergence ratios, 3-D models \textcolor{black}{will be} studied later. One important aspect of the simulations \textcolor{black}{surface} right away \textcolor{black}{for the earlier versions of the design}. A complete halfraum \textcolor{black}{puts} too much plasma in the heating beam path because of more ablated material\textcolor{black}{. Hence, a cone attached to the inner surface of the target is introduced to leave a path for a heating beam while preserving the convergent geometry for field compression.} The \textcolor{black}{subcritical} densities \textcolor{black}{are} either not seen or \textcolor{black}{are} highly dependent on the other parameters in the system when the conical fence \textcolor{black}{is} not included. To keep the setup robust to parameter changes without overcomplicating it, the conical fence \textcolor{black}{is} found to be a solution. Intuitively, more material ablated from a surface causes larger densities in agreement with our simulations. \autoref{fig:setup} shows the cross-section of cone attached to the halfraum target. Note that the cone is a slit fence in 2-D.

Our 2-D simulations \textcolor{black}{use} a computational domain spanning $2.85\,\text{mm}$ with $1.98$ \textmu m resolution square grids, within a conductive box, necessary to preserve physical boundary conditions on the axial field. Note that the wall of the box has been removed from all the plots. \textcolor{black}{This box at the boundaries} does not artificially increase the magnetic field at the location of the sample since the box wall is far removed from the sample location\textcolor{black}{. This can be cross-checked with the} area ratios at the beginning, and the end of the compression \textcolor{black}{being} consistent with the compressed field values \textcolor{black}{returned by the simulations}. Further, the compression beam and the subsequent laser-plasma interactions are not included in the simulation in this paper. While it is important to estimate their impact on the heating beam, our focus is on how far the electron density is from the critical density. Yet, as should be expected, the electron density along the beam path is found to increase as the slit size decreases, and the field strength at full compression tends to diminish as the slit size increases. A slit size of $0.4\,\text{mm}$ \textcolor{black}{is seen to be} sufficient to keep the field strength high while keeping the heating beam path mostly clear of critical density plasma, and the cut for the conical obstacle \textcolor{black}{is} similarly kept at a distance of $0.4\,\text{mm}$ away from the sample location to optimize both density and the field compression. Another simplification inherent to the coil design of \autoref{fig:Blines}, is the constant axial field of $50$ T throughout the initial simulation domain.  \textcolor{black}{PERSEUS do not model how these initial magnetic field lines are provided}, though the values are consistent with what MIFEDS 2.5 can deliver to the coil as shown in \autoref{fig:Blines}. 

\begin{figure*}[!htb]
    \centering
    \begin{subfigure}[t]{0.5\textwidth}
        \centering
        \includegraphics[width=\textwidth]{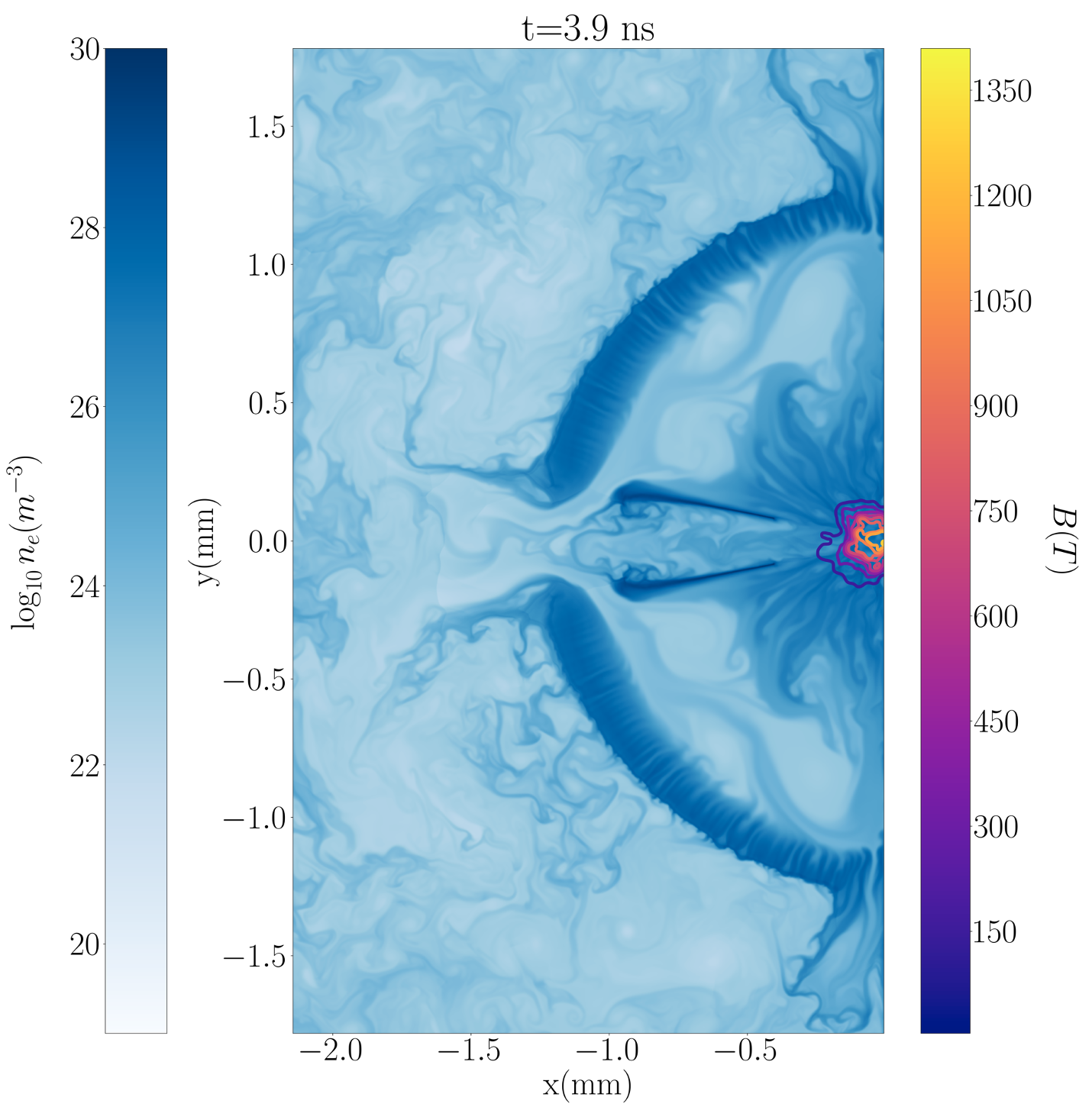}
        \caption{} 
\label{fig:simpeak} 
    \end{subfigure}%
    ~
    \vfill
    \begin{subfigure}[t]{0.6\textwidth}
        \centering
        \includegraphics[width=\textwidth]{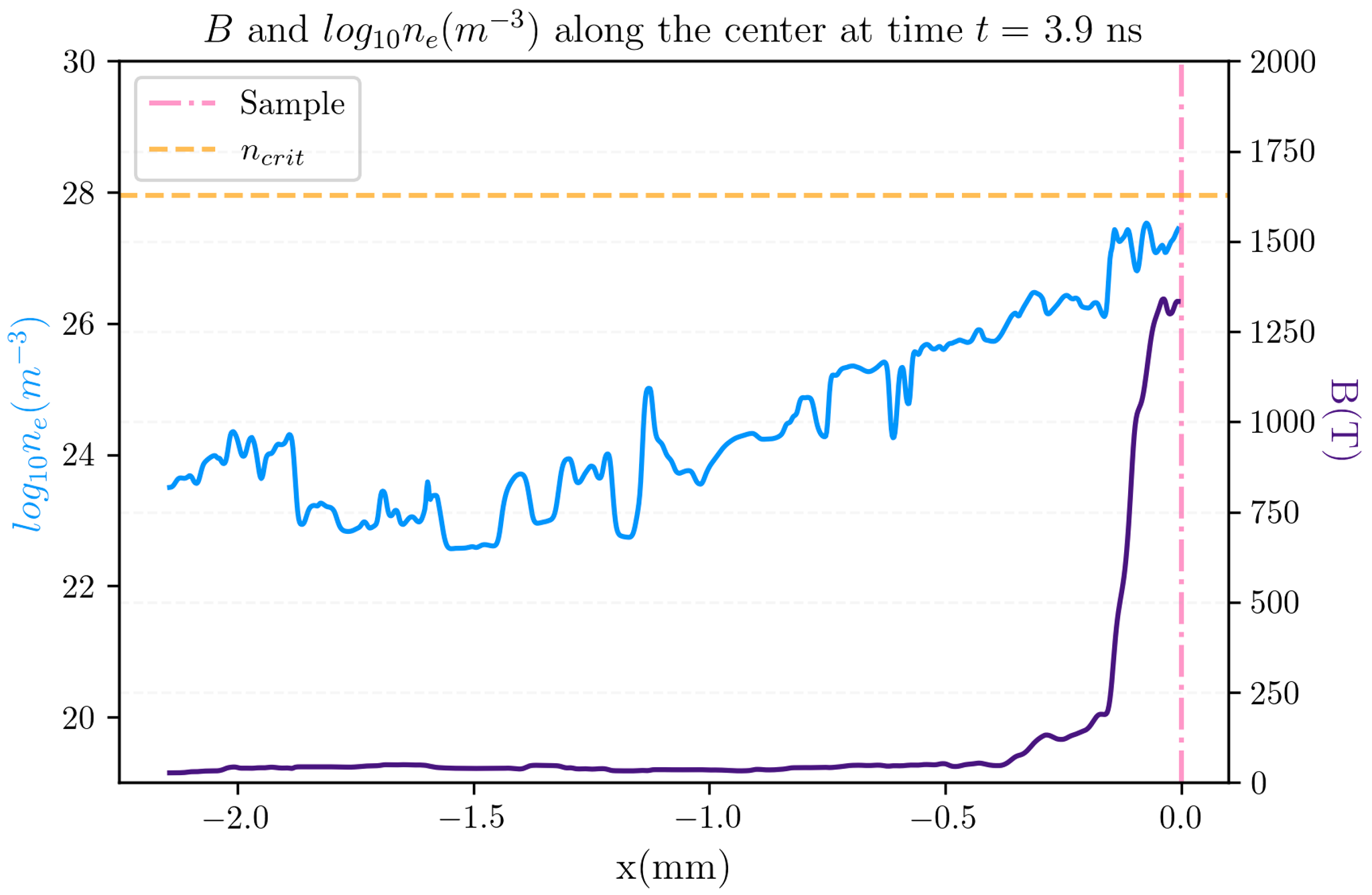}
\caption{} 
\label{fig:lineout} 
\end{subfigure}%
\caption{
        (\subref{fig:simpeak}) $3.9\,\text{ns}$ after the \textcolor{black}{compression} beam hits the halfraum, \textcolor{black}{a} magnetic field strength over $1\,\text{kT}$ is reported by PERSEUS simulations at the sample location. One can also see the ablated electrons' behavior and their distributions on this same plot with the log electron density color bar.
        (\subref{fig:lineout}) The lineout plot right along the x-axis shows subcritical electron densities at the peak field region where the sample will be placed. This plot implies that once successful, our setup would enable introducing more beams to create and diagnose magnetized HED matter samples.
    }
\end{figure*}

The simulation is for the first $10\,\text{ns}$ after the compression beams hit the halfraum. Since plastic targets are commonly used, both the target of halfraum and the sample are modeled to be carbon with an ionization state of 4. For the purposes of this manuscript, however, the field compression is not specific to carbon and could be seen for any other material as well. The laser-plasma interactions here are not simulated, and instead, we provide a power density uniformly distributed on the halfraum's inner surface, corresponding to 20 OMEGA beams, and modeled that as: 
\begin{equation}
P_{d}=A e^{-\left(\frac{r_{in}-r}{\sigma}\right)^2}
\label{energy}
\end{equation}
where $A$ is a power scaling parameter and is \textcolor{black}{$10^{23}\,\text{W/m}^{3}$} for 20 OMEGA beams, and $\sigma$ is the compression beam penetration \textcolor{black}{depth.} $r_{in}=1\,\text{mm}$ is the inner radius of the halfraum and $r$ is the variable for radius centered at \textcolor{black}{the origin}. $\sigma$ was set to four grid cells. Note that in the plots, \autoref{fig:simpeak}\textcolor{black}{, \autoref{fig:lineout}, and   \autoref{fig:nernst}}, there is a shift to have the sample at the origin. All the power density is deposited to the halfraum target. The turn-on time is simulated as a $2\,\text{ns}$ square pulse with a rise time of $0.1\,\text{ns}$.

The simulation results are shown in \autoref{fig:simpeak} and \autoref{fig:lineout} for the time at which $B>1\text{kT}$ is observed along with subcritical densities. \autoref{fig:simpeak} shows the contours for the magnetic field strengths superimposed on the ion density, plotted on the logarithmic scale. We see that $3.9\,\text{ns}$ after the \textcolor{black}{compression} beam hits the target, the maximum magnetic field strength of $1.3\,\text{kT}$ is achieved at the sample location. This can be seen from \autoref{fig:simpeak}. Note that the boundaries are cropped for the plots reported here and the sample is carried to the origin. For \textcolor{black}{$x<-2.5\,\text{mm}$, $y<-2.1375\,\text{mm}$, and $y>2.1375\,\text{mm}$,} the boundary conditions do not allow the flow of particles implying closed \textcolor{black}{boundaries} while for \textcolor{black}{$x>0.35\,\text{mm}$}, so on the right side of the sample, \textcolor{black}{open boundary conditions are introduced} in our simulations. This is because the open boundary conditions are more realistic for the boundary around the sample as closed boundary conditions would unrealistically report even higher field strengths with even higher densities due to the trapping of the particles. Although extended MHD does not assume ideal plasma, plasma still carries the field \textcolor{black}{according to the frozen-in law}. Therefore, for the sample location with field compression, this open boundary avoids such imaginary confinement which would misleadingly report even stronger fields.

\autoref{fig:lineout} \textcolor{black}{is} a lineout taken right at the middle of the y-axis \textcolor{black}{of \autoref{fig:simpeak} so it reflects densities along the line connecting} the slit and the \textcolor{black}{sample.} \textcolor{black}{\autoref{fig:lineout}} clearly shows that the densities between the \textcolor{black}{sample} and the slit are below the critical density which is calculated to be $9\times10^{27}\,\text{m}^{-3}$ for an OMEGA beam of wavelength $351\,\text{nm}$.

\section{\label{sec:discussion}Discussion}


This paper proposes an experimental platform to study magnetized WDM samples. Yet, due to the limited understanding of WDM even the field strengths required to magnetize WDM are not fully known in theory and never studied experimentally. However, it is still obvious based on \textcolor{black}{current understanding that} strong fields above $1$ kT are required for that purpose and these techniques should allow the introduction of another laser to create the WDM \textcolor{black}{state}. Therefore, the design presented here is a first step towards studying magnetized HED \textcolor{black}{matter} and a prototype for possible experiments to study the phase transitions of magnetized WDM samples.

\begin{figure}[!htb]
\centering    
\includegraphics[width=0.5\textwidth]{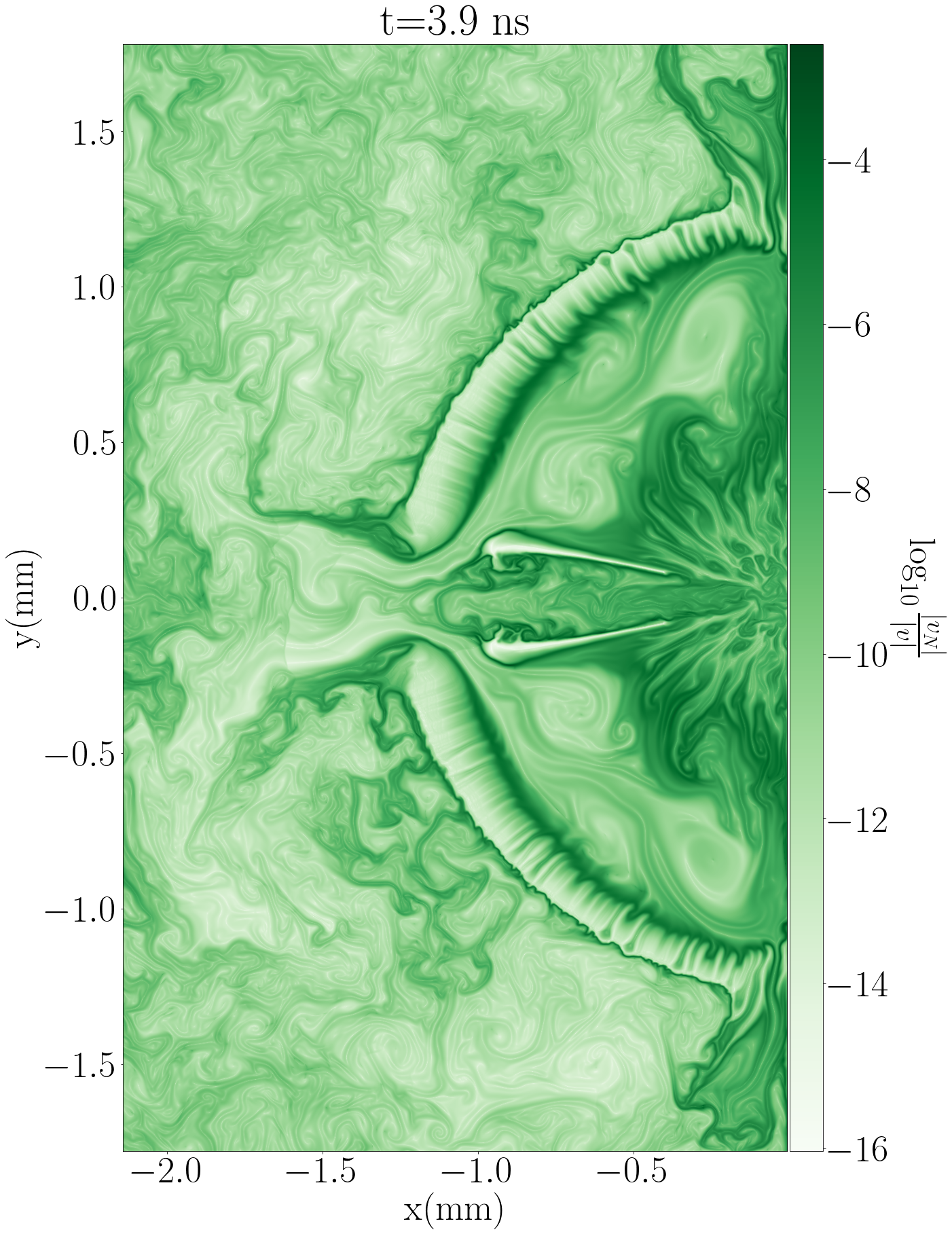}
\caption{\textcolor{black}{The ratio of the calculated magnitude of the Nernst velocity to the flow velocity reported by the code is plotted. Nernst velocity is negligible compared to the flow velocity in the system and therefore, the magnetic field strength should not be reduced due to the Nernst effect.}}
\label{fig:nernst} 
    \end{figure}

The simulations show that strong fields can be generated with low density plasma. However, as in any simulation, there exists room for improvement. One important point is that PERSEUS does not include the Nernst term and having it implemented in our simulations could lead to more realistic results. Yet\textcolor{black}{,} the \textcolor{black}{calculated} Nernst velocities \textcolor{black}{based on simulation outputs} for the \textcolor{black}{time of the} peak field strength indicate that \textcolor{black}{directing the laser to the inner surface of the halfraum target} leads to Nernst velocities to be several orders of magnitude smaller than the flow velocities. In fact, \autoref{fig:nernst} compares the calculated Nernst velocity\cite{10.1063/1.4935286, PhysRevLett.126.075001} with the flow velocity in the simulations. This comparison shows that the Nernst velocity is significantly low and should not impact the reported magnetic field strengths further supporting our simulations.  

In addition to the possible field advection with the Nernst velocity, the magnetic Reynolds number \textcolor{black}{determines the field diffusion}. The simulation data reports magnetic Reynolds numbers larger than unity for the field compression region. This indicates that the frozen-in-flow assumption can be used to describe the plasma behavior under magnetization and the magnetic field should not diffuse away. Therefore, the strong magnetization should be sustained and this further supports our design's capability to achieve field compression.


One possible issue is that the electrons may not be equilibrated at the sample location \textcolor{black}{at} the time \textcolor{black}{when} $B>1\,\text{kT}$ and when the WDM state is created. However, that is not a concern for this work since the simulations do not include the sample and its interaction with the heating beam. The objective is to design an experimental setup, with less emphasis on the specific outcomes that would be obtained from that configuration. 

Although the setup presented here includes a specific set of \textcolor{black}{parameters,} \textcolor{black}{these values can be adjusted}. While testing the feasibility of this design, various simulations with different parameters were run. \textcolor{black}{The setup} is sensitive to changes in halfraum thickness, halfraum radius, slit size, and distance from \textcolor{black}{the} sample to \textcolor{black}{the} cone. Thicker halfraums lead to higher ablated plasma densities, hence causing stronger fields with denser medium. A smaller halfraum radius means easier convergence for the field at the cost of higher density plasma in between, and similarly for the smaller slit size or \textcolor{black}{more} distance from the sample to the cone \textcolor{black}{end}. Denser plasmas would preclude the heating beam to reach the \textcolor{black}{WDM} sample. A feasible set of parameters need to be explored with simulations before performing such an experiment. Yet, the principle of field compression with low density plasma remains robust \textcolor{black}{fundamentally due to the energy conservation.} Therefore, the key parameter for the compression ratio is the laser power deposited\textcolor{black}{. Because the laser energy is} first transferred to plasma formation and kinetic energy, \textcolor{black}{then} through the plasma that energy is converted to magnetic energy.	

Other than geometry, field compression with subcritical plasma is also strongly linked to temperature, density, and laser wavelength. \textcolor{black}{The field compression benefits from having many electrons at high temperature.} While the trade-off between density, and field strength or temperature exists due to the nature of the problem in hand, the principle behind this work is still robust to show that compression ratios similar to \textcolor{black}{existing} field compression setups but with low-density plasmas are achievable for different sets of parameters. We had many other simulations with similar but different geometries and even different laser energies exhibit the same \textcolor{black}{behavior.}

\textcolor{black}{The} ionization state is kept at 4 in the simulations to represent the plastic target used. While this is realistic on average, it \textcolor{black}{can not capture} ionization happening during the simulation \textcolor{black}{like many other MHD codes}. \textcolor{black}{However, our simulations for ionization states of 3 and 5 at the same settings as the one included here still allow subcritical densities along with field strengths above $1$ kT for the same time point. Simulations indicate that the ionization state strongly impacts the field strength. These results from the simulations agree with our intuition based on energy conservation. Energy conservation at full compression can briefly be written between magnetic energy and the thermal energy in the system with no ramp pressure. Setting the plasma frequency to the laser frequency in the energy conservation expression, $\frac{B^2}{2\mu_0}=\left(1+\frac{1}{Z}\right)\frac{\omega_L^{2}m_e \epsilon_0}{e}  T$ is seen with $\omega_L$ being the laser frequency. This expression supports the motivation behind this work that magnetic fields can be compressed with high-temperature plasma. It also indicates that the impact of ionization becomes less important for increasing ionization states. Hence, there is a trade-off between the magnetic field strengths that can be reached and the subcritical densities to be preserved. For the design in hand, the ionization state of 4 remains an optimum but this does not prevent any other ionization state from exhibiting field compression with subcritical density plasma.}

It is demonstrated in this paper that the halfraum setup could be successful in the study of phase transitions in magnetized WDM and open the path to examine magnetized HED matter experimentally. Consequently, the following step would entail the initiation of experiments to further assess the design which should be supported by 3-D simulations.

\section{\label{sec:conc}Conclusion}

In conclusion, an experimental setup for the study \textcolor{black}{of magnetized WDM samples} is presented, and its feasibility is demonstrated through 2-D PERSEUS simulations. The design involves a halfraum configuration, created by bisecting a hohlraum along its axis, to enable PXRDIP \textcolor{black}{diagnostic} for precise examination of the phase transition process. In this geometry, \textcolor{black}{the compression beams are directed to the inner surface of the halfraum target and} the ablated plasma moves inwards, compressing the \textcolor{black}{initial magnetic} field \textcolor{black}{to field strengths} exceeding $1\,\text{kT}$ based on our simulations, enabling the magnetization of WDM's valence electrons. \textcolor{black}{Our} simulation results \textcolor{black}{also} reveal that the plasma density between the sample and the heating beam remains \textcolor{black}{subcritical}, allowing the \textcolor{black}{heating} beam to reach the target. These results show great promise in studying the phase transitions in magnetized WDM, a critical endeavor with significant implications for ongoing research in fusion and \textcolor{black}{astrophysics.}


\begin{acknowledgments}
This material is based upon work supported by the Department of Energy [National Nuclear Security Administration] University of Rochester “National Inertial Confinement Fusion Program” under Award Number DE-NA0004144.

This report was prepared as an account of work sponsored by an agency of the U.S. Government. Neither the U.S. Government nor any agency thereof, nor any of their employees, makes any warranty, express or implied, or assumes any legal liability or responsibility for the accuracy, completeness, or usefulness of any information, apparatus, product, or process disclosed, or represents that its use would not infringe privately owned rights. Reference herein to any specific commercial product, process, or service by trade name, trademark, manufacturer, or otherwise does not necessarily constitute or imply its endorsement, recommendation, or favoring by the U.S. Government or any agency thereof. The views and opinions of authors expressed herein do not necessarily state or reflect those of the U.S. Government or any agency thereof.

The research presented here is supported by NSF grant number PHY-2020249 and the Horton fellowship. We acknowledge the members of the Extreme State Physics laboratory at the University of Rochester and the Innovative Concepts Group in the Experimental Division of LLE for their helpful feedback and discussions. \textcolor{black}{We also thank our reviewers }\textcolor{black}{ for their precious comments that helped us improve this paper.}
\end{acknowledgments}
\section*{Data Availability Statement}
Data available upon request to first author.



\nocite{*}
\bibliography{aipsamp}

\end{document}